\documentclass[prl,twocolumn,superscriptaddress,letterpaper]{revtex4}
\usepackage{graphicx}
\begin{document}
\newtheorem{theorem}{Theorem}
\newtheorem{acknowledgement}[theorem]{Acknowledgment}
\newtheorem{algorithm}[theorem]{Algorithm}
\newtheorem{axiom}[theorem]{Axiom}
\newtheorem{claim}[theorem]{Claim}
\newtheorem{conclusion}[theorem]{Conclusion}
\newtheorem{condition}[theorem]{Condition}
\newtheorem{conjecture}[theorem]{Conjecture}
\newtheorem{corollary}[theorem]{Corollary}
\newtheorem{criterion}[theorem]{Criterion}
\newtheorem{definition}[theorem]{Definition}
\newtheorem{example}[theorem]{Example}
\newtheorem{exercise}[theorem]{Exercise} 
\newtheorem{lemma}[theorem]{Lemma}
\newtheorem{notation}[theorem]{Notation}
\newtheorem{problem}[theorem]{Problem}
\newtheorem{proposition}[theorem]{Proposition}
\newtheorem{remark}[theorem]{Remark}
\newtheorem{solution}[theorem]{Solution}
\newtheorem{summary}[theorem]{Summary}   
\def\r{{\bf{r}}}
\def\j{{\bf{j}}}
\def\m{{\bf{m}}}
\def\k{{\bf{k}}}
\def\kt{{\tilde{\k}}}
\def\mt{{\hat{t}}}
\def\mG{{\hat{G}}}
\def\mg{{\hat{g}}}
\def\mGa{{\hat{\Gamma}}}
\def\mS{{\hat{\Sigma}}}
\def\mT{{\hat{T}}}
\def\K{{\bf{K}}}
\def\P{{\bf{P}}}
\def\q{{\bf{q}}}
\def\Q{{\bf{Q}}}
\def\p{{\bf{p}}}
\def\x{{\bf{x}}}
\def\X{{\bf{X}}}
\def\Y{{\bf{Y}}}
\def\F{{\bf{F}}}
\def\G{{\bf{G}}}
\def\bG{{\bar{G}}}
\def\mbG{{\hat{\bar{G}}}}
\def\M{{\bf{M}}}
\def\V{\cal V}
\def\tchi{\tilde{\chi}}
\def\tx{\tilde{\bf{x}}}
\def\tk{\tilde{\bf{k}}}
\def\tK{\tilde{\bf{K}}}
\def\tq{\tilde{\bf{q}}}
\def\tQ{\tilde{\bf{Q}}}
\def\si{\sigma}
\def\ep{\epsilon}
\def\hep{{\hat{\epsilon}}}
\def\al{\alpha}
\def\be{\beta}
\def\ep{\epsilon}
\def\bep{\bar{\epsilon}_\K}
\def\up{\uparrow}
\def\de{\delta}
\def\De{\Delta}
\def\up{\uparrow}
\def\dwn{\downarrow}
\def\ksi{\xi}
\def\etha{\eta}
\def\product{\prod}
\def\goto{\rightarrow}
\def\switch{\leftrightarrow}
\title{Evidence for Excimer Photoexcitations in an Ordered $\pi$-Conjugated Polymer Film}
\author{K.~Aryanpour} 
\affiliation{Department of Physics, University of Arizona, Tucson, Arizona 85721, USA}

\author{C.-X.~Sheng}
\affiliation{Department of Physics, University of Utah, Salt Lake City, Utah 84112, USA}
\affiliation{School of Electronic and Optical Engineering, Nanjing University of Science and
Technology, Nanjing, Jiangsu, China 210094}

\author{E.~Olejnik} 
\affiliation{Department of Physics, University of Utah, Salt Lake City, Utah 84112, USA}

\author{B.~Pandit}
\affiliation{Department of Physics, University of Utah, Salt Lake City, Utah 84112, USA} 

\author{D.~Psiachos}
\affiliation{ICAMS, Ruhr-Universitaet Bochum, Bochum 44801, Germany}

\author{S.~Mazumdar} 
\affiliation{Department of Physics, University of Arizona, Tucson, Arizona 85721, USA}
\affiliation{College of Optical Sciences, University of Arizona, Tucson, Arizona 85721, USA}

\author{Z.~V.~Vardeny} 
\affiliation{Department of Physics, University of Utah, Salt Lake City, Utah 84112, USA}
\date{\today}
\begin{abstract}
We report pressure-dependent transient picosecond and continuous wave photomodulation studies of disordered and ordered films of 2-methoxy-5-(2-ethylhexyloxy) poly(para-phenylenevinylene) (MEH-PPV). 
Photoinduced absorption (PA) bands in the disordered film exhibit very weak pressure-dependence and are assigned to intrachain excitons and polarons. In contrast, the ordered film exhibits two additional transient PA bands in the mid-infrared that blueshift dramatically with pressure. Based on high-order configuration interaction calculations we ascribe the PA bands in the ordered film to excimers. Our work brings new insight to the exciton binding energy in ordered films versus disordered films and solutions.

\end{abstract}
\maketitle 
\par Ordered $\pi$-conjugated polymer films exhibit photophysics remarkably different from dilute solutions or disordered films \cite{Rothberg06a,Arkhipov04a,Schwartz03a,Conwell01a}. Explanations given for these differences include photoexcitation branching into intrachain excitons and polarons in the ordered films \cite{Miranda01a}, as well as formation of a variety of intermolecular species \cite{Rothberg06a,Arkhipov04a,Schwartz03a,Conwell01a,Jenekhe94a,Webster96a}. The distinctive behavior of the ordered films are due to strong interchain interaction absent in dilute solutions or disordered films. It follows that the ability to vary the extent of interchain interactions in a controlled manner would provide an ideal tool for understanding the role of morphology on the photophysics in these materials. Here we report such a study: we probe pressure effects on the transient picosecond (ps) and continuous wave (cw) PM spectra of disordered and ordered MEH-PPV films up to $119$ kbar. The ordered film exhibits two correlated PA bands missing in the disordered films, which dramatically blue-shift with pressure. We further show, both experimentally and theoretically, that this key experimental result cannot be explained within scenarios involving exciton delocalization, or photogeneration of bound polaron-pairs. Our calculations establish unambiguously that the primary photoexcitated species in ordered MEH-PPV films are excimers, whose PA bands are expected to show pressure-induced blueshift. Our wavefunction analysis of initial and final states of PA bands also gives physical understanding behind the
reduced exciton binding energy in ordered films \cite{Alvarado98a}.
\par We used thin films of polymer samples drop-cast on quartz substrates from powder as received from ADS. Transient photomodulation (PM) spectroscopy was utilized to resolve the primary photoexcitations. Specifically, we used femtosecond (fs) two-color pump-probe correlation technique with a low-power (energy/pulse $\sim 0.1$ nJ), high repetition rate ($\sim 80$ MHz) laser system based on a Ti:sapphire (Tsunami, Spectra-Physics) laser having a temporal pulse resolution of $150$ fs \cite{Sheng07a}. The pump $\hbar\omega$ was frequency doubled to $\hbar\omega = 3.1$ eV; the output beam of an optical parametric oscillator, OPO (Opal, Spectra-Physics) was used as a probe with $\hbar\omega$ ranging from $0.24$ to $1.1$ eV \cite{Sheng07a}. The pump and probe beams were focused on the film surface inside the cryostat or diamond anvil pressure cell to a spot $\sim 50$ $\mu$m in diameter; and the focus area on the film was frequently changed to prevent photoinduced fatigue. The transient PM signal was measured using a phase sensitive lock-in technique at modulation frequency 30 kHz provided by an acousto-optic modulator. In the mid-IR spectral range we only obtained photoinduced absorption (PA), which is given as the fractional change in transmission  
$\Delta T/T(t)$. Both $T$ and $\Delta T$ were measured using solid state photodetectors Ge, InSb and MCT depending on the spectral range. For the steady-state PM spectra we employed a cw laser for pump excitation and an incandescent light as a probe \cite{Jiang02a}, using a standard PM set-up based on a $\frac{1}{4}$ met monochromator or FTIR spectrometer.
\begin{figure} 
\includegraphics[width=2.7in]{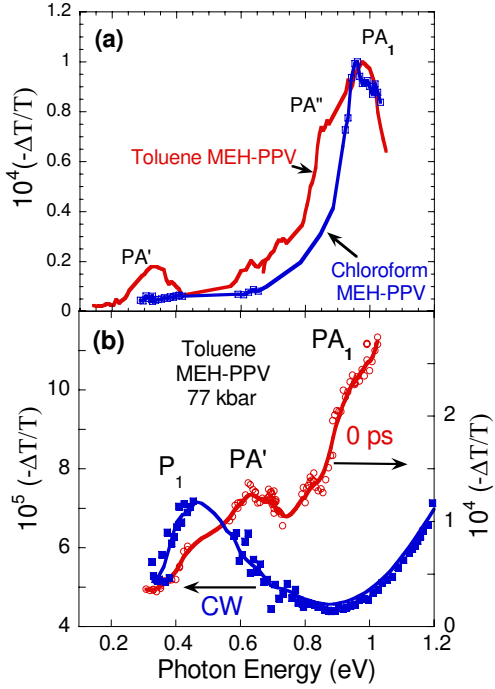}
\caption[a]{(Color online) (a) Transient ps PM spectra at $t=0$ of toluene-formed (red) and chloroform-formed (blue) MEH-PPV films. The PA exciton band, PA$_1$, and excimer bands PA$^{'}$ and PA$^{''}$ are assigned.
(b) Transient (red) and cw (blue) PM spectra of toluene-formed MEH-PPV at hydrostatic pressure of $77$ kbar. PA$_1$ and PA$^{'}$ are as in (a); P$_1$ is a polaron band.}
\label{exp1} 
\end{figure}
\par Fig.~\ref{exp1}a shows the transient PM spectra at time $t=0$ of two MEH-PPV films cast from toluene and chloroform solutions, respectively. Toluene is a poor solvent for MEH-PPV \cite{Rothberg06a,Schwartz03a}, and the films cast from its solution contain coexisting ordered and disordered phases \cite{Rothberg06a}. MEH-PPV cast from chloroform solution, in contrast, is dominated by the disordered phase \cite{Rothberg06a,Schwartz03a}. The transient PM of the chloroform-formed MEH-PPV contains a single PA band, PA$_1$ peaked at $\sim 0.95$ eV, which is very similar to PA$_1$ of MEH-PPV in solution and is correlated to the stimulated emission band in the visible spectral range \cite{Sheng07a}. We therefore identify it as due to intrachain excitons. The toluene-formed MEH-PPV shows two {\it additional} PA bands, PA$^{'}$ at $\sim 0.35$ eV and a shoulder PA$^{''}$ at $\sim 0.85$ eV. We assign these bands to interchain species formed in the ordered phase.
\par
\begin{figure*} 
\includegraphics[width=6.4in]{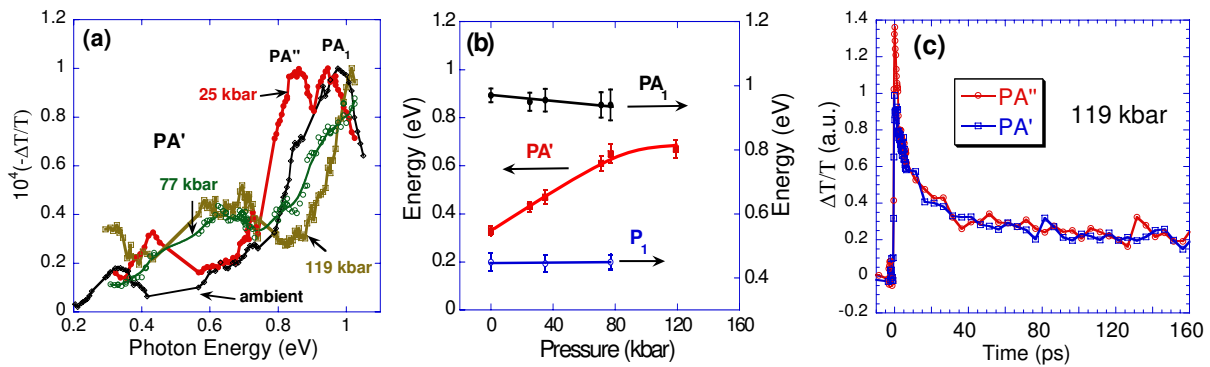}
\caption[a]{(Color online) (a) ps transient ($t=0$) PM spectra of toluene-formed MEH-PPV at ambient pressure, and {\it P =} $25$, $77$, and $119$ kbar; the PA bands are assigned as in Fig.\ref{exp1}. (b) Summary of the peak positions for the transient PA$^{'}$ (red) and PA$_{1}$ (black) at $t=0$; and cw P$_{1}$ (blue) vs. {\it P}. (c) The decay dynamics of the transient PA$^{'}$ and PA$^{''}$ bands at $P=119$ kbar.}
\label{exp2} 
\end{figure*}

 The band PA$^{'}$ in MEH-PPV was identified previously as due to polarons \cite{Sheng07a}, since the cw polaron band P$_{1}$ also peaks at about the same $\hbar\omega$(probe) = 0.4 eV \cite{Drori07a}. To confirm our current assignment of PA$^{'}$ to interchain species we applied high hydrostatic pressure {\it P} to the toluene-formed MEH-PPV film in a pressure cell. The MEH-PPV film was peeled off the substrate under a microscope, and placed in a diamond-anvil cell equipped with IR-transmitting windows, that was filled with a pressure-transmitting liquid, perfluoro-tri-butylamine to ensure hydrostatic pressure. The pressure inside the cell was measured via the pressure-induced blue-shift in the polymer IR-active C-H frequency at $\sim 3000$ $\textrm{cm}^{-1}$, which was pre-calibrated against the pressure-induced change of the well known PL lines of a ruby chip. Fig.\ref{exp1}b shows the transient and cw PM spectra under pressure $P=77$ kbar. PA$^{'}$ blue-shifts substantially to $\sim 0.65$ eV, in contrast to the polaron P$_{1}$ band observed in the cw PM spectrum, which does not shift with pressure (Fig.\ref{exp2}b). PA$^{'}$ is therefore not due to polarons.
\par Fig.~\ref{exp2}a shows the MEH-PPV transient PM spectrum for various increasing {\it P}-values. PA$^{'}$ and PA$^{''}$ both grow in intensity relative to PA$_{1}$, and also blue shift together, whereas PA$_{1}$ remains at $\sim 0.95$ eV. The energy shifts of PA$_{1}$, PA$^{'}$ and cw P$_1$ due to polarons are plotted vs. {\it P} up to $120$ kbar in Fig.~\ref{exp2}b. PA$^{'}$ shifts by about $0.35$ eV up to $P=77$ kbar, and peaks at the same $\hbar\omega$(probe) at still higher {\it P}; in contrast, $P_{1}$ and PA$_{1}$ do not shift much with pressure. Fig.~\ref{exp2}c shows the decay dynamics of PA$^{'}$ and PA$^{''}$ at $P = 119$ kbar. The dynamics are clearly identical, coming from the same photoexcitation species. 
\par To understand the origin of the new species in the ordered film we model the ordered phase as two interacting, cofacially stacked, planar PPV oligomers of equal length \cite{Wang08a,Huang08a}. Previously we have shown that high-order configuration interaction (CI) calculations inclusive of all quadruple excitations are essential for finding the PA$^{'}$ band \cite{Wang08a}. This necessitates the use of the semiempirical $\pi$-electron Hamiltonian and also limits the lengths of our oligomers to 3 units. Even with such small system size for the two-chain system our basis size is 1.8 million. For computational simplicity we chose a symmetric arrangement, where each carbon atom of one oligomer lies directly on top of the equivalent carbon atom of the second oligomer. Our calculations are based on the Hamiltonian $H =\sum_{\mu=1,2} H_{\mu}+H_{\mu,\mu^{'}}$, where $H_{\mu}$ is the single-chain Pariser-Parr-Pople Hamiltonian \cite{Pariser53a}, 
\begin{eqnarray}
\label{eq:H_intra}
H_{\mu} = -\sum_{\langle ij \rangle, \sigma}t_{ij}
(c_{\mu,i,\sigma}^\dagger c_{\mu,j,\sigma}+ H.C.) + \nonumber \\
 \sum_{i} Un_{\mu,i,\uparrow}n_{\mu,i,\downarrow}+\sum_{i<j} V_{ij} (n_{\mu,i}-1)(n_{\mu,j}-1).
\end{eqnarray}
\par Here $c_{\mu,i,\sigma}^{\dagger}$ creates a $\pi$-electron of spin $\sigma$ on carbon atom $i$ of the $\mu$th molecule, $n_{i,\mu,\sigma} =  c_{\mu,i,\sigma}^{\dagger}c_{\mu,i,\sigma}$ and $n_{\mu,i} = \sum_{\sigma}n_{\mu,i,\sigma}$. We chose the nearest neighbor hopping matrix element $t_{ij}=t=2.4$ eV for phenyl C-C bonds, and $2.2$ ($2.6$) eV for the intrachain single (double) C-C bonds \cite{Chandross97a}. U is the repulsion between two electrons occupying the same $p_{z}$ orbital of a C atom, and $V_{ij}$ are intrachain intersite Coulomb interactions parametrized as $V_{i,j} = U/(\kappa\sqrt{1+0.6117 R_{ij}^2})$, where $R_{ij}$ is the distance between C atoms i and j in Angstroms. We chose $U=8.0$ eV and $\kappa=2$ \cite{Chandross97a}.
\par The intermolecular Hamiltonian is written as \cite{Wang08a},
\begin{eqnarray}
\label{eq:H_inter}
H_{\mu,\mu'} = -\sum_{\langle ij \rangle, \sigma}t^{\perp}_{ij}
(c_{\mu,i,\sigma}^\dagger c_{\mu',j,\sigma}+ H.C.) + \nonumber \\
 \sum_{i<j} V^{\perp}_{ij} (n_{\mu,i}-1)(n_{\mu',j}-1).
\end{eqnarray}
where $t^{\perp}_{ij}=t^{\perp}$ is restricted to nearest neighbors. For $V^{\perp}_{ij}$ we chose the same functional form as the intrachain $V_{ij}$ defined above, with a screening parameter $\kappa^{\perp}\le\kappa$ \cite{Aryanpour10a}. We report calculations for $\kappa^{\perp}=\kappa=2$. We modeled the effects of increasing pressure by decreasing the intermolecular distance from $0.41$ nm at ambient pressure to $0.37$ nm at the highest pressures, while increasing $t^{\perp}$ from $0.07$ eV to $0.15$ eV. Our goal is to understand effects of enhanced pressure at a qualitative level only, which in turn allows determination of the dominant interchain species in the ordered phase. Besides reducing intermolecular distances, pressure also causes planarizarion of individual chains \cite{Schmidtke07a}, that (a) increases the effective conjugation length, which only decreases intramolecular transition energies, and (b) enhances $t^{\perp}$, an effect we have included. Our basis set consists of the Hartree-Fock orbitals of the individual molecules. The localized basis allows calculation of the total charge on the individual oligomers (hereafter ionicity $\rho$), for each eigenstate. 
\begin{figure*}
\includegraphics[width=6.7in]{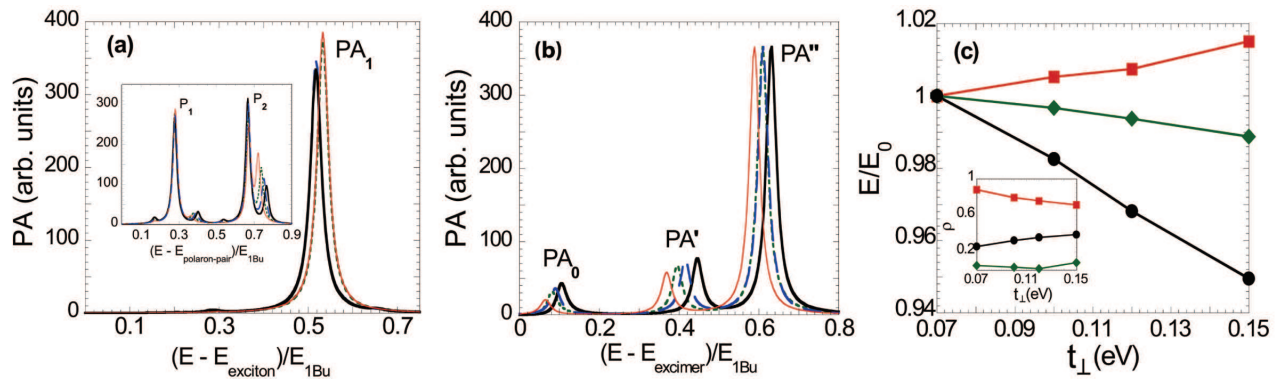}
\caption[a]{(Color online) 
(a) Calculated dependence of PA bands from (a) the two-chain exciton and (b) the excimer, 
on intermolecular interactions. Red thin curves
correspond to $t^{\perp}=0.07$ eV, intermolecular separation $d = 0.41$ nm; green dotted curves to $t^{\perp}=0.1$ eV, $d = 0.4$ nm; blue dashed curves to $t^{\perp}=0.12$ eV, $d = 0.38$ nm; and black thick cruves to $t^{\perp}=0.15$ eV, $d = 0.37$ nm. The inset in (a) shows the dependence of PA bands from the polaron-pair on $t^{\perp}$. (c) Energies (E) of the excimer (black circles), final state of PA$^{'}$ (red squares) and the final state of PA$^{''}$ (green diamonds) as a function of $t^{\perp}$, normalized by their energies E$_{0}$ at the $t^{\perp}=0.07$ eV (ambient pressure). The inset shows the corresponding ionicities. The $t^{\perp}$ and matching $d$ are all the same as in (a) and (b).}
\label{theo1} 
\end{figure*}

\par The intermolecular species we investigated are: (i) the delocalized covalent ($\rho=0$) optical exciton $\big|{exc_1}\big> + \big|{exc_2}\big>$, where $\big|exc_{j}\big>$ implies excitation on the jth molecule; (ii) the completely ionic ($\rho=1$) Coulombically bound polaron-pair,  $\big|P_{\mu}^{+}P_{\mu^{\prime}}^{-}\big>$, where $P_{\mu}^{+}(P_{\mu^{\prime}}^{-})$ is a positively (negatively) charged oligomer \cite{Rothberg06a,Arkhipov04a,Conwell01a}; and (iii) the excimer or charge-transfer exciton \cite{Wang08a,Jenekhe94a,Webster96a,Marciniak07a}, $\big|CTX\big>$, 
\begin{eqnarray}
\label{eq:excimer}
\big|\textrm{CTX}\big> = c_{c}\big(\big|{exc_1}\big> - \big|{exc_2}\big>\big)  + c_{i}\big(\big|P_1^+P_2^-\big> - \nonumber \\ \big|P_2^+P_1^-\big>\big)  + \cdots  
\end{eqnarray}
where $\cdots$ denotes higher-order terms; for the $\big|CTX\big>$, $0<\rho<1$. To obtain PA from the polaron-pair state forming below the optical exciton, we create donor-acceptor PPV oligomers by adding $\sum_{\mu,i}(-1)^{\mu}\epsilon n_{\mu,i}$  to  $H_{\mu}$ in Eq.\ref{eq:H_intra}, and also reduce $\kappa^{\perp}$. For $\epsilon=0.185$ and $\kappa^{\perp}=1.3$ the lowest excited state is the polaron-pair with $\rho=0.9$ \cite{Aryanpour10a}. 
\par In Fig.~\ref{theo1}a we show our calculated PA$_{1}$ band of the optical exciton belonging to the two-chain system, as a function of increasing $t^{\perp}$. Here and in Fig.~\ref{theo1}b we scaled all excitation energies by the energy E$_{1B_u}$ of the single chain exciton. As readily seen, PA$_{1}$ exhibits a weak redshift at the highest $t^{\perp}$. Our calculated bands from the polaron-pair state are shown in the in the inset of Fig.~\ref{theo1}a; increased intermolecular interaction does not affect the PA. We therefore conclude that PA$^{'}$ and PA$^{''}$ bands in the ps transient PM spectra of the ordered films cannot be ascribed to excitons or polaron-pairs. 
\par The weak pressure effect on the optical exciton and polaron-pair PA is due to their extreme ionicities, $\rho=0$ and 1, respectively, that do not change with pressure. In contrast, increased intermolecular interactions enhances $c_i$ in Eq.~\ref{eq:excimer}, which, in turn, leads to shifts in the PA energies. In Fig.~\ref{theo1}b we show the calculated PA bands from $\big|CTX\big>$, which is the lowest excited state of our Hamiltonian, again for several different $H_{\mu,\mu^{'}}$. The lowest energy PA$_{0}$ lies outside our experimental spectral range \cite{Wang08a}, whereas PA$^{'}$ and PA$^{''}$ are both seen in the experiment. PA$^{'}$ (PA$^{''}$) originates predominantly from the polaron-pair (exciton) component of  $\big|CTX\big>$. The pressure-induced blueshifts of the experimental PA$^{'}$ and PA$^{''}$ energies shown in Fig.~\ref{exp2}a are replicated in Fig.~\ref{theo1}b. Although quantitative comparisons are difficult, with $E_{1Bu}\sim 2.2$ eV for MEH-PPV the scaled energy shift of the PA$^{'}$ band at the largest $H_{\mu,\mu^{'}}$ 
is about $0.25$ eV, which is close to the maximum measured blueshift of this band shown in Fig.\ref{exp2}a ($\sim$ 0.3 eV).
\par In Fig.~\ref{theo1}c we give a mechanistic explanation for the observed blueshifts of the excimer PA bands under pressure. Increased $H_{\mu,\mu^{'}}$  leads to larger $\rho$ for the  $\big|CTX\big>$ and significant decrease in its energy. Our calculated $\rho$ for the final state of PA$^{'}$ is considerably larger than $\rho$ for $\big|CTX\big>$ (see inset), indicating that the PA$^{'}$ absorption, over and above intramolecular polaronlike excitations has also  strong contribution from intermolecular charge-transfer excitation. Indeed, the decrease (increase) in energy of the initial (final) state of $PA^{'}$, accompanied by the increase (decrease) in ionicity of the corresponding wavefunction is the classic signature of enhanced intermolecular charge-transfer. In contrast to $\rho$ for the final state of PA$^{'}$, $\rho$ for the final state of PA$^{''}$ transition (see inset) is very small. This state is thus reached from the neutral exciton component of $\big|CTX\big>$. The increase in PA$^{''}$ peak energy comes mostly from the decrease in $\big|CTX\big>$ energy with {\it P}. 
\par Our analysis in Fig.~3c gives insight to smaller exciton binding energies in ordered films than in solutions of $\pi$-conjugated polymers \cite{Alvarado98a}. The final state of PA$_1$ is to the mA$_g$ two-photon state, which has greater {\it intrachain} charge-transfer character than the 1B$_u$ optical exciton \cite{Chandross99a}. The much larger interchain charge-transfer character of the final state of PA$^{'}$ than in the excimer shows that it is the {\it exact interchain equivalent of the mA$_g$.} 
Smaller exciton binding energy in ordered films \cite{Alvarado98a} is thus not due to
screening of Coulomb interactions \cite{vanderHorst99a}, but to the appearance of new low lying states with greater charge separation
\cite{Ruini02a}. 
\par In summary, the significant pressure induced blueshifts exhibited by the transient PA bands in ordered MEH-PPV films indicate 
{\it intermediate ionicity for the primary photoexcitation,} {\it i.e.,} an excimer. 
Understanding the role of morphology in determining the photophysical behaviour of $\pi$-conjugated polymer 
films is crucial for their applications in next generation optoelectronic devices. Our joint theory-experiment
work gives a {\it new diagnostic tool} for determining the nature of the primary photoexcitations
in polymer films, and we expect this and similar techniques to have wide applications.
\par The Utah group thanks Valentina Morandi and Josh Holt for help with the high pressure and ps measurements. The work at Utah was supported in part by the NSF grant DMR-0803325. The Arizona group thanks Alok Shukla and Zhendong Wang for computational help. The work at Arizona was partially supported by NSF grant DMR-0705163. C.-X.S thanks the support of National Natural Science Foundation of China grant No.61006014.

\begin{thebibliography}{21}
%
\bibitem{Rothberg06a}
L.~Rothberg, {\it Photophysics of Conjugated Polymers, Semiconducting Polymers: Chemistry, Physics and Engineering}, Vol.{\bf I}, edited by G.~Hadziioannou and G.~G.~Malliaras (John Wiley, 2006), pp. 179-204.
%
\bibitem{Arkhipov04a}
V.~I.~Arkhipov, and H.~Bassler, Phys.\ Stat.\ Sol. (a) {\bf 201}, 1152-1187 (2004).
%
\bibitem{Schwartz03a}
B.~J.~Schwartz, Annu.\ Rev.\ Phys.\ Chem. {\bf 54}, 141-172 (2003).
%
\bibitem{Conwell01a}
E.~M.~Conwell, {\it Photophysics of conducting polymers, in Organic Electronic Materials: Conjugated Polymers and Low Molecular Weight Solids}, edited by R. Farchioni and G. Grosso (Springer, New York, 2001), pp. 127-180.
%
\bibitem{Miranda01a}
P.~B.~Miranda, D.~ Moses, and A.~J.~Heeger, Phys.\ Rev.\ B {\bf 64}, 081201 (2001).
%
\bibitem{Jenekhe94a}
S.~A.~Jenekhe, and J.~A.~Osaheni, Science {\bf 265}, 765-768 (1994).
%
\bibitem{Webster96a}
S.~Webster, and D.~N.~Batchelder, Polymer {\bf 37}, 4961-4968 (1996).
%
\bibitem{Alvarado98a} S. F. Alvarado {\it et al.}, Phys. Rev. Lett. {\bf 81}, 1082 (1998).
%
\bibitem{Sheng07a}
C.-X.~Sheng, M.~Tong, and Z.~V.~Vardeny, Phys.\ Rev.\ B. {\bf 75}, 085206 (2007).
%
\bibitem{Jiang02a}
X.~M.~Jiang {\it et al.}, Adv.\ Funct.\ Materials {\bf 12}, 587- 594 (2002).
%
\bibitem{Drori07a}
T.~Drori {\it et al.}, Phys.\ Rev.\ B. {\bf 76}, 033203 (2007).
%
\bibitem{Wang08a}
Z.~Wang, S.~Mazumdar, and A.~Shukla, Phys.\ Rev.\ B {\bf 78}, 235109 (2008).
%
D.~Psiachos, and S.~Mazumdar, Phys.\ Rev.\ B {\bf 79}, 155106 (2009).
%
\bibitem{Huang08a}
Y.-S.~Huang {\it et al.}, Nat.\ Mater. {\bf 7}, 483-489 (2008).
%
\bibitem{Pariser53a}
R.~Pariser, and R.~G.~Parr, J.\ Chem.\ Phys. {\bf 21},  9865-9867 (1953).
%
J.~A.~Pople, Trans. Faraday Soc. {\bf 49}, 767-776 (1953).
%
\bibitem{Chandross97a}
M.~Chandross {\it et al.}, Phys.\ Rev.\ B, {\bf 55}, 1486-1496 (1997).
%
\bibitem{Aryanpour10a}
K.~Aryanpour, D.~Psiachos, and S.~Mazumdar, Phys.\ Rev.\ B {\bf 81}, 085407 (2010).
%
\bibitem{Schmidtke07a} J. P. Schmidtke {\it et al.}, Phys. Rev. Lett. {\bf 99}, 167401 (2007).
%
\bibitem{Marciniak07a}
H.~Marciniak {\it et al.}, Phys.\ Rev.\ Lett. {\bf 99}, 176402 (2007).
%
\bibitem{Chandross99a} M. Chandross, Y. Shimoi, and S. Mazumdar, Phys. Rev. B {\bf 59},
4822 (1999)

\bibitem{vanderHorst99a} J.-W. van der Horst {\it et al.} Phys. Rev. Lett. {\bf 83}, 4413 (1999).

\bibitem{Ruini02a} A. Ruini {\it et al.} Phys. Rev. Lett. {\bf 88}, 206403 (2002).
\end{thebibliography}
\end{document}